\documentclass[]{article}
\usepackage{epsf,amsmath,amssymb}
\begin{document}

\title{$M00N$ states in ion-laser interactions}
\author{A. Z\'u\~niga-Segundo$^1$, R. Ju\'arez-Amaro$^2$,\\
F. Soto-Eguibar$^3$ and H. M. Moya-Cessa$^3$}
\date{Today}
\maketitle

\noindent {$^1$ Departamento de F\'{\i}sica, Escuela Superior de F\'{\i}sica y Matem\'aticas Edificio 9, Unidad Profesional 'Adolfo L\'opez Mateos', Mexico\\
$^2$ Universidad Tecnol\'ogica de la Mixteca, Apdo. Postal 71, Huajuapan de Le\'on, Oax., 69000 Mexico\\
$^3$ Instituto Nacional de Astrof\'{\i}sica, \'Optica y Electr\'onica, Calle Luis Enrique Erro No. 1, 72840 Santa Mar\'{\i}a Tonantzintla, Mexico}

\begin{abstract} A new class of entangled states, similar to N00N states is introduced. We call these states M00N states as the excitations shared in both subsystems do not need to be equal. The generation proposed here does not need conditional measurements, and therefore is achieved in a deterministic manner.

\end{abstract}

\section{Introduction}

In recent years, the generation of nonclassical states has attracted a great deal of attention. Among most preferred
states because of their nonclassical behaviour are (a) macroscopic quantum superpositions of quasiclassical coherent states with different mean phases or amplitudes \cite{1,2}, (b) squeezed states \cite{3}, (c) the particularly important limit of extreme squeezing; i.e., Fock or number states, and more recently, (d) nonclassical states of combined photon pairs also called N00N states \cite{4,5}. N00N states, because of their entanglement properties, are particularly useful in the quantum information area.

Methods to measure generated non-classical states have already been presented. For instance, by looking at the atomic inversion, squeezed states \cite{6} and superposition of coherent states \cite{7} show specific signatures of their statistical properties.

It is well known that N00N states can be used to obtain high-precision phase measurements, becoming more and more advantageous as the number of photons grows. Many applications in quantum imaging, quantum information and quantum metrology \cite{9} depend on the availability of entangled photon pairs because entanglement is a distinctive feature of quantum mechanics that lies at the core of many new applications. These maximally path-entangled multiphoton states may
be written in the form
\begin{equation}
\mid N00N\rangle_{ab}=\frac{1}{\sqrt{2}}\left(\mid N\rangle_a\mid 0\rangle_b+\mid 0\rangle_a\mid N\rangle_b\right)\;.\label{eq01}
\end{equation}
It has been pointed out that $N00N$ states manifest unique coherence properties by showing that they exhibit a periodic transition between spatially bunched and antibunched states when undergo Bloch oscillations. The period of the bunching/antibunching oscillation is $N$ times faster than the period of the oscillation of the photon density \cite{10}.

The greatest $N$ for which $N00N$ states have been produced is $N=5$ \cite{4}. Most schemes to generate this class of states are either for optical \cite{4,5} or microwave \cite{11} fields.

In this contribution we will show how to generate a new class of states, namely, $M00N$ states, or states of the form
\begin{equation}
\mid M00N\rangle_{ab}=\frac{1}{\sqrt{2}}\left(\mid M\rangle_a\mid 0\rangle_b+\mid 0\rangle_a\mid N\rangle_b\right)\;.\label{eq02}
\end{equation}
with high excitation numbers $M$ and $N$.

\section{Ion vibrating in two dimensions}

We consider an ion in a two-dimensional Paul trap \cite{11}, and we assume that the ion is driven by a plane wave
\begin{equation}
E^{(-)}(\hat{x},\hat{y},t)=E_0\hbox{e}^{-i(k_x\hat{x}+k_y\hat{y}+\omega t)}\;,\label{eq03}
\end{equation}
with $k_j$, $j= x,y$ the wavevectors of the driving field. The Hamiltonian has the form
\begin{multline}
H=\nu_x\hat{a}^\dagger_x\hat{a}_x+\nu_y\hat{a}^\dagger_y\hat{a}_y+\frac{\omega_{21}}{2}\hat{\sigma}_z+\Omega_x\left\{\hbox{e}^{-i\left[\eta_x(\hat{a}_x+\hat{a}^\dagger_x)+\omega t\right]}\hat{\sigma}_++H.C.\right\}+\\+\Omega_y\left\{\hbox{e}^{-i\left[\eta_y(\hat{a}_y+\hat{a}^\dagger_y)+\omega t\right]}\hat{\sigma}_++H.C.\right\}\;,\label{eq04}
\end{multline}
where we have defined the Lamb-Dicke parameters
\begin{equation}
\eta_x=2\pi\frac{\sqrt{_x\langle 0\mid\Delta\hat{x}^2\mid 0\rangle_x}}{\lambda_x}\;,\quad
\eta_y=2\pi\frac{\sqrt{_y\langle 0\mid\Delta\hat{y}^2\mid 0\rangle_y}}{\lambda_y}\;,
\label{eq05}
\end{equation}
and redefined the ladder operators according to
\begin{equation}
k_x\hat{x}=\eta_x(\hat{a}_x+\hat{a}^\dagger_x)\;,\quad k_y\hat{y}=\eta_y(\hat{a}_y+\hat{a}^\dagger_y)\;.\label{eq06}
\end{equation}
In the resolved sideband limit, the vibrational frequencies $n_x$ and $n_y$ are much larger than other characteristic
frequencies and the interaction of the ion with the two lasers can be treated separately using a nonlinear
Hamiltonian \cite{17,18,15}. 

We consider that the ion is trapped in the $x$-axis; i.e., $\Omega_x\neq 0$ and $\Omega_y=0$; then
\begin{equation}
H_x=\nu_x\hat{a}^\dagger_x\hat{a}_x+\frac{\omega_{21}}{2}\hat{\sigma}_z+\Omega_x\left\{\hbox{e}^{-i\left[\eta_x(\hat{a}_x+\hat{a}^\dagger_x)+\omega t\right]}\hat{\sigma}_++H.C.\right\}\;.\label{eq07}
\end{equation}
We write $\omega_{21}=\omega-\delta$, where $\delta$ is the detuning, to obtain
\begin{equation}
H_x=\nu_x\hat{a}^\dagger_x\hat{a}_x+\frac{\omega+\delta}{2}\hat{\sigma}_z+\Omega_x\left\{\hbox{e}^{-i\left[\eta_x(\hat{a}_x+\hat{a}^\dagger_x)+\omega t\right]}\hat{\sigma}_++H.C.\right\}\;.\label{eq08}
\end{equation}
We transform the Hamiltonian to a frame rotating at $\omega$ frequency by means of the transformation
\begin{equation}
T=\hbox{e}^{-i\frac{\omega t}{2}\hat{\sigma}_z}\;,\label{eq09}
\end{equation}
and we get
\begin{equation}
H_x=\nu_x\hat{a}^\dagger_x\hat{a}_x+\frac{\delta}{2}\hat{\sigma}_z+\Omega_x\left\{\hbox{e}^{-i\left[\eta_x(\hat{a}_x+\hat{a}^\dagger_x)\right]}\hat{\sigma}_++\hbox{e}^{i\left[\eta_x(\hat{a}_x+\hat{a}^\dagger_x)\right]}\hat{\sigma}_-\right\}\;.\label{eq10}
\end{equation}
Using the Baker-Hausdorff formula \cite{19}, and expanding the exponentials in Taylor series, we cast the Hamiltonian
to
\begin{equation}
H_x=\nu_x\hat{a}^\dagger_x\hat{a}_x+\frac{\delta}{2}\hat{\sigma}_z+\Omega_x\left[\hbox{e}^{-\frac{\eta^2_x}{2}}\sum_{n,m}
\frac{{(-i\eta_x)}^n}{n!}\frac{{(-i\eta_x)}^m}{m!}\hat{a}^{\dagger n}_x\hat{a}^m_x\hat{\sigma}_++H.C.
\right]\;.\label{eq11}
\end{equation}
Transforming to the interaction picture,
\begin{equation}
H_{Ix}=\Omega_x\hbox{e}^{-\frac{\eta^2_x}{2}}\sum_{n,m}\frac{{(-i\eta_x)}^n}{n!}\frac{{(-i\eta_x)}^m}{m!}\hat{a}^{\dagger n}_x\hat{a}^m_x\hat{\sigma}_+\hbox{e}^{i\eta_x(n-m+k)}+H.C.
\;.\label{eq12}
\end{equation}
We consider now the low-intensity regime; i.e., $\Omega_x\ll\eta_x$, and we apply the rotating wave approximation, to get
\begin{equation}
H_{Ix}=\Omega_x\hbox{e}^{-\frac{\eta^2_x}{2}}(-i\eta_x)^k
\sum_{n=0}^\infty\frac{{(-\eta_x)}^{2n}}{n!(k+n)!}\hat{a}^{\dagger n}_x\hat{a}^{k+n}_x\hat{\sigma}_++H.C.
\;,\label{eq13}
\end{equation}
by substituting $\hat{a}^{\dagger n}_x\hat{a}^n_x=\frac{\hat{n}!}{(\hat{n}-n)!}$, multiplying by $\frac {(\hat{n}+k)!}{(\hat{n}+k)!}$, and rearranging terms
$$H_{Ix}=\Omega_x\hbox{e}^{-\frac{\eta^2_x}{2}}(-i\eta_x)^k \frac{\hat{n}!}{(\hat{n}+k)!}L^k_{\hat n}(\eta^2_x)\hat{a}^k_x\hat{\sigma}_++H.C.\;,$$
where we have identified
$L^k_{\hat n}(\eta^2_x)=\sum^{\hat n}_{n=0}\frac{(-1)^n(\eta^2_x)^n}{n!}\frac{(\hat{n}+k)!}{(n+k)!(\hat{n}-n)!}$, with the associated Laguerre polynomials; so that finally
\begin{equation}
H_{Ix}=\Omega_x\left(f^k_x(\hat{n})\hat{a}^k_x\hat{\sigma}_+\hat{a}^{\dagger k}_xf^{*k}_x(\hat{n})\hat{\sigma}_-\right)\;,\label{eq14}
\end{equation}
where
\begin{equation}
f^k_x(\hat{n})=\hbox{e}^{-\frac{\eta^2_x}{2}}(-i\eta_x)^k \frac{\hat{n}!}{(\hat{n}+k)!}L^k_{\hat n}(\eta^2_x)\;.\label{eq15}
\end{equation}
By using a nonunitary transformation in terms of Susskind-Glogower phase operator \cite{20} (written here in
matrix form) we write
\begin{equation}
H_{Ix}=\begin{pmatrix} 
 1 & 0 \\
 0 & \hat{V}^{\dagger k}_x
\end{pmatrix}
H_{Ix}\begin{pmatrix}
 1 & 0 \\
 0 & \hat{V}^{k}_x
\end{pmatrix}\;,\label{eq16}
\end{equation}
where we have introduced and defined
\begin{equation}
H_{Ix}=\Omega_x f^k_x(\hat{n})\sqrt{\hat{a}^k_x\hat{a}^{\dagger k}_x}\hat{\sigma}_++\Omega_x f^{*k}_x(\hat{n})\sqrt{\hat{a}^k_x\hat{a}^{\dagger k}_x}\hat{\sigma}_-
\;.\label{eq17}
\end{equation}
The evolution operator for this last transformed Hamiltonian, $U_{Ix}=\hbox{e}^{-iH_{Ix}t}$, may be calculated easily. For
this we need
\begin{equation}
H^{2m}_{Ix}=\Omega^{2m}_x\mid f^k_x(\hat{n})\mid^{2m}\left(\sqrt{\hat{a}^k_x\hat{a}^{\dagger k}_x}\right)^{2m}I_{2\times 2}\;,\label{eq18}
\end{equation}
where $I_{2\times 2}$ is the $(2\times 2)$ unity matrix. For odd powers we have
\begin{multline}
H^{2m+1}_{Ix}=\hat{\sigma}_+\Omega^{2m+1}_x\frac{\mid f^k_x(\hat{n})\mid^{2m+1}{f^{*k}_x(\hat{n})}}{\mid f^k_x(\hat{n})\mid}\left(\sqrt{\hat{a}^k_x\hat{a}^{\dagger k}_x}\right)^{2m+1}+\\
+\hat{\sigma}_-\Omega^{2m+1}_x\frac{\mid f^k_x(\hat{n})\mid^{2m+1}{f^{*k}_x(\hat{n})}}{\mid f^k_x(\hat{n})\mid}\left(\sqrt{\hat{a}^k_x\hat{a}^{\dagger k}_x}\right)^{2m+1}
\;.\label{eq19}
\end{multline}
Then
\begin{equation}
U_{Ix}(t)=\sum_m\frac{(-it)^{2m}}{(2m)!}H^{2m}_{Ix}+ \sum_m\frac{(-it)^{2m+1}}{(2m+1)!}H^{2m+1}_{Ix}\;,\label{eq20}
\end{equation}
and therefore
\begin{equation}
U_{Ix}(t)=\begin{pmatrix}
 C_{ee} & S_{eg} \\
 S_{ge} & C_{gg}
\end{pmatrix}\;,\label{eq21}
\end{equation}
with
\begin{align}
C_{ee}&=\cos\left(\Omega_xt\mid f^k_x(\hat n)\mid\sqrt{\hat{a}^k_x\hat{a}^{\dagger k}_x}\right)\;,\label{eq22}\\
S_{eg}&=-i(-i)^k\sin\left(\Omega_xt\mid f^k_x(\hat n)\mid\sqrt{\hat{a}^k_x\hat{a}^{\dagger k}_x}\right)\hat{V}^k_x\;,\label{eq23}\\
S_{ge}&=-i\hat{V}^{\dagger k}_x(i)^k\sin\left(\Omega_xt\mid f^k_x(\hat n)\mid\sqrt{\hat{a}^k_x\hat{a}^{\dagger k}_x}\right)\;,\label{eq24}\\
\nonumber
\end{align}
and
\begin{equation}
C_{gg}=\hat{V}^{\dagger k}_x\cos\left(\Omega_xt\mid f^k_x(\hat n)\mid\sqrt{\hat{a}^k_x\hat{a}^{\dagger k}_x}\right)\hat{V}^{k}_x\;.\label{eq25}\\
\end{equation}
Finally, as $\sqrt{\hat{a}^k_x\hat{a}^{\dagger k}_x}=\sqrt{\frac{(\hat{n}+k)!}{\hat{n}!}}$, we can write
\begin{equation}
U_{Ix}(t)=\begin{pmatrix}
 U_{11} & U_{12} \\
 U_{21} & U_{22}
\end{pmatrix}\;,\label{eq26}
\end{equation}
with
\begin{align}
U_{11}&=\cos\left[\Omega_xt\mid f^k_x(\hat n)\mid\sqrt{\frac{(\hat{n}+k)!}{\hat{n}!}}\right]\;,\label{eq27}\\
U_{12}&=(-i)^{k+1}\sin\left[\Omega_xt\mid f^k_x(\hat n)\mid\sqrt{\frac{(\hat{n}+k)!}{\hat{n}!}}\right]\hat{V}^k_x\;,\label{eq28}\\
U_{21}&=-(i)^{k+1}\hat{V}^{\dagger k}_x\sin\left[\Omega_xt\mid f^k_x(\hat n)\mid\sqrt{\frac{(\hat{n}+k)!}{\hat{n}!}}\right]\;,\label{eq29}\\
\nonumber
\end{align}
and
\begin{equation}
U_{22}=\hat{V}^{\dagger k}_x\cos\left[\Omega_xt\mid f^k_x(\hat n)\mid\sqrt{\frac{(\hat{n}+k)!}{\hat{n}!}}\right]\hat{V}^{k}_x\;.\label{eq30}\\
\end{equation}

Now, we consider as initial state of the ion a number state $\mid n\rangle$ for the vibrational motion and the excited state $\mid e\rangle$ for the internal states; i.e.,
\begin{equation}
\mid\psi(0)\rangle=\begin{pmatrix} \mid n\rangle\\ 0\\
\end{pmatrix}\;.\label{eq31}
\end{equation}
The probability, after the time $t$, of finding the ion in its internal excited state is then
\begin{multline}
P_e(t)=\sum^{\infty}_{m=0}\langle e\mid\langle m\mid\psi(t)\rangle\langle\psi(t)\mid m\rangle\mid e\rangle\\\times\cos^2\left(\Omega_xt\mid f^k_x(\hat n)\mid\sqrt{\frac{(\hat{n}+k)!}{\hat{n}!}}\right)
\;.\label{eq32}
\end{multline}
It is clear that after a time
\begin{equation}
t_0=\frac{\pi}{2}\frac{1}{\Omega_x\mid f^k_x(\hat n)\mid}\sqrt{\frac{n!}{(n+k)!}}
\;,\label{eq33}
\end{equation}
the probability to find the ion in its internal excited state is $0$, so at that time the ion is in its internal ground state with probability $1$. This situation is obviously repeated periodically; every $2j+1$, $j=0,1,2,\cdots$ times $t_0$, the ion will be in its ground state. As can bee seen from Hamiltonian (\ref{eq17}), when the probability of finding the ion
in the excited state goes to zero, the ion is giving four phonons to the vibrational motion. Now, if we consider the ion initially in its ground state, the probability to find it in the ground state at the same time t0 is also zero. In this case the ion removes four phonons of the vibrational motion. 

If we consider now the interaction only in the $y$-axis; i.e., $\Omega_y\neq 0$ and $\Omega_x=0$, we get exactly the same expressions and the same results with the variable $y$ instead of $x$.

\section{Generation of M00N states}

By starting with the ion in the excited state and the vibrational state in the vacuum state; i.e., $\mid 0\rangle_x\mid 0\rangle_y$, if we set $\eta_y=0$, after the time $t_0$ when the probability to find the ion in its excited state is zero (meaning that the ion, by passing from its excited to its ground state, gives 4 phonons to the vibrational motion), we can generate the state $\mid 4\rangle_x\mid 0\rangle_y$. Repeating this procedure (with the ion reset again to the excited state, via a rotation), but now with $\eta_x=0$, four phonons are added to the $y$-vibrational motion, generating the two-dimensional state $\mid 4\rangle_x\mid 4\rangle_y$.

Therefore, if we consider the ion initially in a superposition of ground and excited states, and the $\mid 4\rangle_x\mid 4\rangle_y$ vibrational state; i.e.,
\begin{equation}
\mid\psi_{init}\rangle=\frac{1}{\sqrt{2}}(\mid e\rangle+\mid g\rangle)\mid 4\rangle_x\mid 4\rangle_y\;,\label{eq34}
\end{equation}
for $\eta_y=0$ and $t_0$, the state generated is
\begin{equation}
\mid\psi_{\eta_y=0}\rangle=\frac{i}{\sqrt{2}}(\mid e\rangle\mid 0\rangle_x+\mid g\rangle\mid 8\rangle_x)\mid 4\rangle_y\;.\label{35}
\end{equation}
Now, we consider this state as initial state for the next interaction with $\eta_x=0$ and still the interaction time $t_0$, to produce
\begin{equation}
\mid\psi_{\eta_x=0}\rangle=-\frac{1}{\sqrt{2}}(\mid e\rangle\mid 0\rangle_x\mid 8\rangle_y+\mid g\rangle\mid 8\rangle_x\mid 0\rangle_y)\;.\label{eq36}
\end{equation}
If in equation (\ref{eq17}) we consider $k=2$, the vibration in the $y$ axis and $\eta_y\ll 1$ we obtain the Hamiltonian
\begin{equation}
H^{(2)}_{Iy}=\Omega_y(\hat{a}^2_y\hat{\sigma}_++\hat{a}^{\dagger 2}_y\hat{\sigma}_-)\;.\label{eq37}
\end{equation}
It is known that this interaction is periodic and may be used to subtract/add excitations from the system \cite{21}. We use this fact, and by setting the interaction time such that, if the ion is initially in its excited state, after a time tg it will end up in its ground state, if the initial state is given by (\ref{eq36}), the state produced is
\begin{equation}
\mid\psi\rangle=-\frac{1}{\sqrt{2}}\mid g\rangle(\mid 8\rangle_x\mid 0\rangle_y +\mid 0\rangle_x\mid 10\rangle_y)\;,\label{eq38}
\end{equation}
this is, the part of the entangled state in (\ref{eq36}) associated with the excited states ``wins" to excitations, while the one associated to the ground state remains invariant, leaving, without conditional measurement, a state we have named
$M00N$ state.

\section{Conclusions}

We have shown how $M00N$ states for the two dimensional vibrational motion of an ion in a Paul trap may be generated without the need of conditional measurements. By means of a set of laser interactions, the ion is manipulated in such a way that excitations may be added or subtracted in a controlled form in order to produce the target state.

\end{document}